\documentclass[aip,graphicx,reprint]{revtex4-1}

\usepackage{graphicx}
\usepackage{dcolumn}
\usepackage{bm}
\usepackage{color}
\usepackage{amsmath}
\usepackage{mathtools}
\usepackage{booktabs} 
\usepackage[T1]{fontenc}
\usepackage[ngerman,english]{babel}
\usepackage{amsfonts}
\usepackage{amsmath}
\usepackage{array}
\usepackage{epsf}
\usepackage{epsfig}
\usepackage{float}
\usepackage[version=3]{mhchem}
\usepackage{textcomp}
\usepackage{graphics}
\usepackage{epstopdf}
\usepackage{placeins}
\usepackage{footnote}
\usepackage{bbm}
\usepackage{enumerate}

\newcommand{\Fe}{[Fe(H$_2$O)$_{6}$]$^{2+}$}

\begin{document}

\title{Ultrafast spin-flip dynamics in transition metal complexes triggered by soft X-ray light} 

\author{Huihui Wang}
\author{Sergey I. Bokarev}
\email{sergey.bokarev@uni-rostock.de}
\affiliation{ Institut f\"{u}r Physik, Universit\"{a}t Rostock,
  Albert-Einstein-Str. 23-24, 18059 Rostock, Germany}
\author{Saadullah G. Aziz}
\affiliation{Chemistry Department, Faculty of Science, King Abdulaziz University, 21589
  Jeddah, Saudi Arabia}
\author{Oliver K\"{u}hn}
\affiliation{ Institut f\"{u}r Physik, Universit\"{a}t Rostock,
  Albert-Einstein-Str. 23-24, 18059 Rostock, Germany}

\date{\today}

\begin{abstract}
Recent advances in attosecond physics provide access to the correlated motion of valence and core electrons on their intrinsic  timescales.  For valence excitations,  processes related to the  electron spin are usually driven by nuclear motion.  For core-excited states, where the core hole has a nonzero angular momentum, spin-orbit coupling is strong enough to drive spin-flips on a much shorter time scale. Here, unprecedented short spin-crossover driven by spin-orbit coupling is demonstrated  for L-edge (2p$\rightarrow$3d) excited states of a prototypical Fe(II) complex. It occurs on a time scale, which is   faster than the core hole lifetime of about 4~fs. A detailed analysis of such phenomena will help to gain a fundamental understanding of spin-crossover processes and build up the basis for their control by light.
\end{abstract}


\maketitle 
The rapid development of gas-phase and surface high harmonic generation techniques paves the way to study ultrafast processes occurring in the  soft X-ray domain~\cite{Dromey2006,Zhang2015,Popmintchev2015} and on ultrashort time scales approaching few tens of attoseconds.~\cite{Teichmann2016}  The novel light sources provide a tool to measure, trigger, and control ultrafast electronic processes in atoms, molecules, and nanoparticles for both valence and more tightly bound core electrons via preparation of intricate superpositions of quantum states.~\cite{Kling2008,Ivanov2009,Lepine2013}   Attosecond spectroscopy has a huge potential to study atomic and molecular responses to incident light,~\cite{Picon_2016,Healion_2012}  thus giving access to, for example, electron correlation manifesting itself in the entanglement of bound- and photo-electrons (shake-ups), Auger and interatomic Coulomb decay, as well as to the coupling of electrons in plasmonic systems.~\cite{Kling2008,Ivanov2009,Sansone2012,Lepine2013} Further,  progress has been seen for the understanding of the dynamics of  charge (hole) migration~ in molecules driven solely by electron correlation.~\cite{Hennig2005,Remacle2006,Sansone2012,Kuleff2014,Nisoli2014,Kraus2015} Microscopic understanding of such ultrafast transfer phenomena  is essential, e.g., to approach the fundamental limits of the transmission speed of electronic signals relevant for molecular electronics. 

Devices based on spin-polarized currents are a prospective extension of conventional electronics.~\cite{Dietl_Spintronics_2008,Bader2010}  Recently, spin-crossover dynamics attracted much attention, e.g., in the context of high-density magnetic data storage devices.~\cite{Cannizzo2010,Halcrow2013}   Popular materials are based on  Fe(II) organometallic complexes. Due to their partially filled 3d-shell they feature  low- as well as high-spin electronic states. Upon valence excitation these systems exhibit an ultrafast spin-crossover occurring on time scales of the order of 100~fs.~\cite{Auboeck2015} The spin-orbit coupling (SOC) constants for valence excited states, however, are small and  spin-crossover is essentially driven by nuclear motion.  Thus the time scale is determined by the related vibrational periods (see also Ref.~\citenum{Bargheer_2002}).  For core-excited electronic states of transition metal complexes, however, the magnitude of SOC increases dramatically.  Therefore, one expects the spin dynamics  to change from a nuclear to an electronically driven process.  In this Letter, it will be demonstrated for a prototypical Fe(II) coordination compound that electronically driven spin-crossover after core hole excitation indeed takes place on a few femtoseconds time scale.

\begin{figure*}[t]
\includegraphics{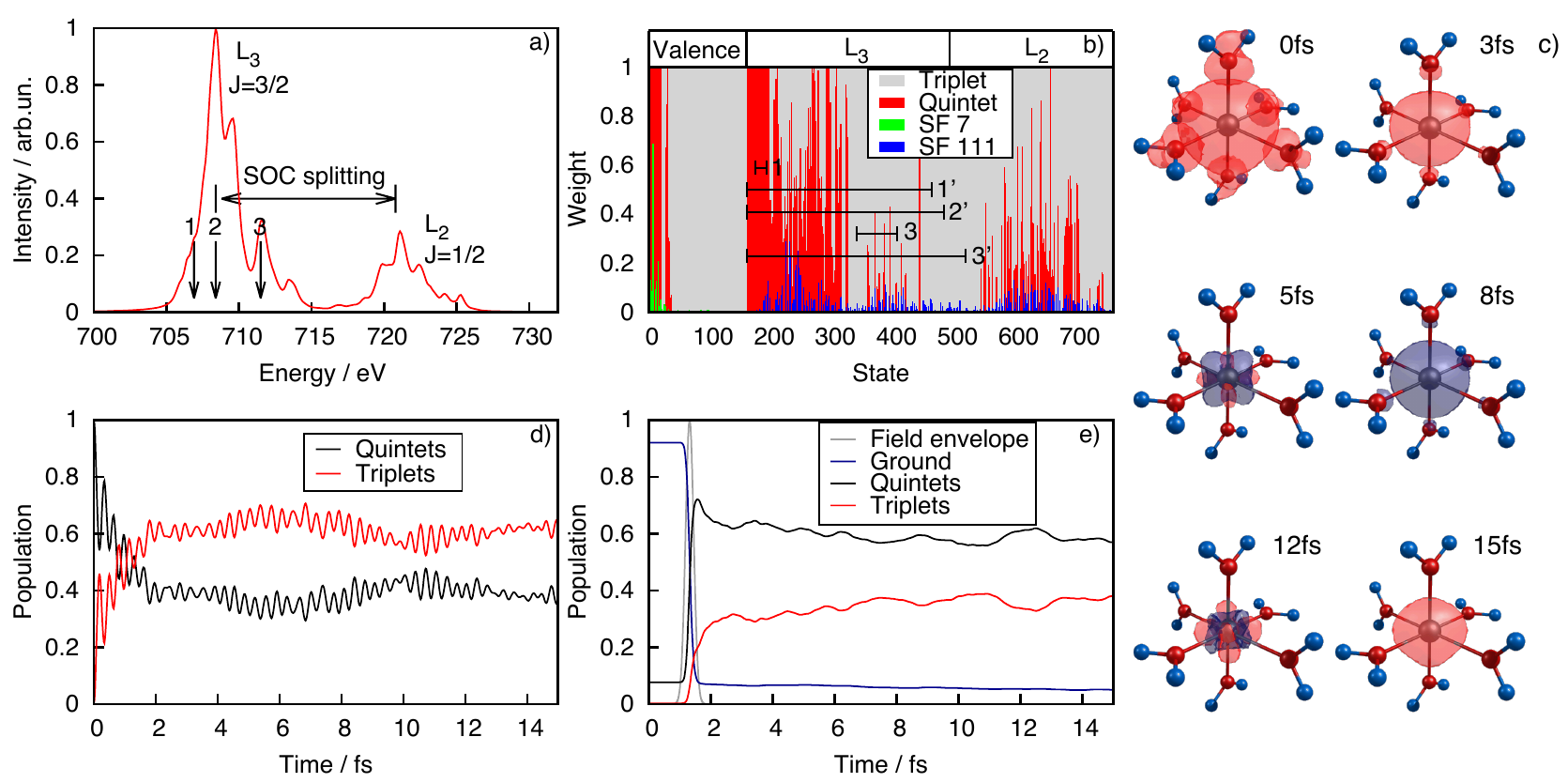}
\caption{
a) X-ray absorption spectrum of \Fe; arrows denote the excitation energies considered in case II. 
b) Contributions of the quintet (red) and triplet (grey) SF to the SOC states. The particular contributions of  SF7 (green bars) and SF111 (blue bars) to the different SOC states are also shown. Numbered ranges 1 and 3 as well as 1', 2', and 3' correspond to the states excited by pulse of 0.5\,eV and 5.0\,eV widths, respectively, for the excitation energies marked in panel a). 
c) Evolution of spin-density difference ($\rho_{\uparrow}-\rho_{\downarrow}$) obtained in case I for a an initial state, which  corresponds to the SF111  with $M_S=+2$.
d) Evolution of the total population of the quintet and triplet electronic states after instantaneous excitation to the SF111 state (case I). 
e) Evolution of the total population of the quintet and triplet electronic states after  explicit field excitation (case II) with the pulse centered at $\hbar \Omega=708.4$~eV having width of $\hbar/\sigma=5.0$~eV, $t_0=$1.32\,fs, and $E_0=$2.5\,$\rm E_he^{-1}bohr^{-1}$.
The envelope of the excitation pulse is shown in grey. The population of $M_S$-components of the ground state is shown by the blue line. }
\label{Fig_soc}
\end{figure*} 

Ultrafast spin-flip is investigated using the Time-Dependent Restricted Active Space Configuration Interaction (TD-RASCI) method, which is similar in spirit to the techniques proposed in Refs.~\citenum{Tremblay2011,Kato2012}. 
Here, the  electronic wave function within clamped nuclei approximation is represented in the basis of Configuration State Functions (CSFs), $\Phi_j^{(S,M_S)}$, with the total spin $S$ and its projection $M_S$:
%
\begin{equation}\label{WF}
|\Psi (t)\rangle =\sum_{j}{c_{j}^{(S,M_S)}(t) |\Phi_j^{(S,M_S)}}\rangle \, .
\end{equation}
%
The CSFs are obtained  using a fixed molecular orbital basis, optimized at the restricted active space self-consistent field~\cite{Malmqvist1990} level prior to propagation. The Hamiltonian in the CSF basis reads
%
\begin{eqnarray}
	\label{Ham}
   \mathbf{H}(t)&= &\mathbf{H}_{\rm CI}+\mathbf{V}_{\rm SOC}+\mathbf{U}_{\rm ext}(t) \nonumber \\
   &=&
   \left(
   \begin{array}{cc}\nonumber 
      \mathbf{H}_{h}  &  0 \\
      0  & \mathbf{H}_{l} \\
   \end{array}
   \right)
   +   
   \left(
   \begin{array}{cc} 
      \mathbf{V}_{hh}  &  \mathbf{V}_{hl} \\
      \mathbf{V}_{lh}  &  \mathbf{V}_{ll} \\
   \end{array}
   \right) 
   +
  \left(
  \begin{array}{cc} 
      \mathbf{U}_{h}(t) &  0 \\
      0  &  \mathbf{U}_{l}(t) \\
   \end{array}
   \right)\\
\end{eqnarray}
where we separated blocks of low ($l$) and high ($h$) spin states. In Eq.~\eqref{Ham}, $\mathbf{H}_{\rm CI}$ is the configuration interaction (CI) Hamiltonian containing the effect of electron correlation. The eigenstates of $\mathbf{H}_{\rm CI}$ will be called spin-free (SF) states. SOC is contained in $\mathbf{V}_{\rm SOC}$, whose matrix elements are calculated within a perturbative LS-coupling scheme,~\cite{AMFI1996} which has demonstrated good performance for  L-edge spectra of transition metal compounds.~\cite{Josefsson2012,Bokarev2013,Chergui2014}  It   provides an intuitive interpretation in terms of pure spin-states with well-defined $S$ and $M_S$ quantum numbers. The eigenstates of $\mathbf{H}_{\rm CI}+\mathbf{V}_{\rm SOC}$ are called SOC states. The interaction with the time-dependent electric field, $\mathbf{U}_{i}=-\vec{\mathbf{d}}_{ii} \cdot \vec{E}(t)$, is taken in a semi-classical dipole approximation with the transition dipole matrices $\vec{\mathbf{d}}_{ii}$ and  the field vector, $\vec{E}(t)$, having a carrier frequency $\Omega$ and a Gaussian envelope, i.e.\ $E(t)=E_0\cos(\Omega t)\exp(-(t-t_0)^2/(2\sigma^2))$. The resulting time-dependent Schr\"odinger equation  has been solved with a 4th order  Runge-Kutta method and using adaptive step size control. 

In the following, the outlined approach is applied to the spin-flip dynamics in \Fe, whose X-ray absorption and resonant inelastic X-ray scattering characteristics has been studied in Refs.~\citenum{Bokarev2013,Atak2013,Golnak2016}. The active space used in TD-RASCI calculations contains 12 electrons distributed over the three 2p (one hole allowed) and five 3d (full CI) orbitals to describe the core excited electronic states corresponding to the dipole allowed 2p$\rightarrow$3d transitions.~\cite{Bokarev2013,Atak2013,Grell2015,Golnak2016}  This active space includes up to 4h4p configurations and results in 35 quintet ($S=2$) and 195 triplet ($S=1$) electronic states, directly interacting via SOC according to the $\Delta S=0,\pm1$ selection rule.  Accounting for the different $M_S$ components, the total amount of the SF and SOC  states is 760, where 160 are valence and 600 core ones. Evaluation of the matrix elements of $\mathbf{H}_i$, $\mathbf{V}_{ij}$, and $\vec{\mathbf{d}}_{ii}$ in the CSF basis has been performed with a locally modified MOLCAS 8.0~\cite{Aquilante2016} quantum chemistry package, applying the relativistic ANO-RCC-TZVP basis set~\cite{Basis1,Basis2} for all atoms. 

In Fig.~\ref{Fig_soc}a) the L-edge absorption spectrum of \Fe is shown for further reference.  It has a shape characteristic for transition metals, featuring  the L$_3$ ($J=3/2$) and L$_2$ ($J=1/2$) bands split due to the  SOC. This splitting is 12.7~eV (SOC constant is 8.5\,eV) what corresponds to a timescale of about 0.33~fs. Panel b) of Fig.~\ref{Fig_soc} illustrates the degree of spin-mixing for the SOC states. It can be seen that the valence excited states are mostly pure quintets or triplets. In contrast the core excited states are dominantly spin-mixtures. 

The dynamics discussed below is driven solely by electronic SOC, while nuclei are fixed at the ground state equilibrium positions. To justify the use of  the clamped nuclei approximation, we assume that the  system is excited far from conical intersections and that the considered time interval  is shorter than the relevant vibrational periods. For \Fe, the  Fe--O stretching and O--Fe--O deformation modes possibly influencing the  2p$\rightarrow$3d core excited electronic states have periods above~100~fs. 

In the following we will discuss the two different cases, illustrating the dynamics of ultrafast spin-crossover. In \textit{case I}, $\vec{E}(t)=0$ and it is assumed  that a particular SF state has been prepared. This somewhat artificial initial condition will serve as a reference, which highlights the spin dynamics driven solely by SOC. In \textit{case II} the system is initially in the ground state with the $M_S$-components of the lowest closely lying electronic states being populated according to the Boltzmann distribution at 300~K. The core hole is created and thus spin dynamics is driven by an ultrashort X-ray pulse, $\vec{E}(t)$, which is  linearly polarized along the shortest of the Fe--O bonds. The spectral overlaps of the different pulses with the SOC states are shown in Fig.~\ref{Fig_soc}b). Note that the field strengths, despite of their large magnitudes, at soft X-ray wavelengths correspond to the weak field regime with Keldysh parameter $\gamma >  7$. Moreover, the transition dipoles are quite small and the Rabi energy with respect to the strongest transition is $d_{\rm max}E_0=$ 2.7~eV and 1.6~eV for the broad and  for the narrow pulse, respectively. In fact,  $E_0$ has been chosen merely to have an appreciable depletion of the ground state for illustration purposes. Indeed, the dynamics triggered by much weaker pulses  coincides qualitatively with the present one; see Supplement.

\textit{Case I.} We have chosen two quintet SF states, i.e.\ number 7 and 111, as initial states for investigating the SOC-driven spin dynamics; for the contributions of  SF7 and SF111 to the SOC states, see  Fig.~\ref{Fig_soc}b). SF7 is a superposition of valence excited SOC states. It turns out that it features a rather weak SOC, such that there is little dynamics within the considered time window of 15~fs (see Supplement). Hence, it will not be discussed further. 
In contrast, SF111, which corresponds to   $M_S=+2$ (four spin-up electrons) has contributions of SOC states from essentially the whole core hole excited L$_3$ and L$_2$ bands. 

Panels c) and d) of Fig.~\ref{Fig_soc} show snapshots of the time evolution of the spin-density difference, $\rho_{\uparrow}-\rho_{\downarrow}$,   and  the total populations of all quintet and triplet states, respectively. A more detailed analysis in terms of the different $M_S$-components  is given in the Supplement. As a consequence of SOC, the  population spreads over both quintet and triplet states such that the total triplet population becomes even larger than the corresponding quintet one within about 1~fs (Fig.~\ref{Fig_soc}d)). The  population transfer occurs according to $\Delta M_S=0,\pm1$ selection rule within and between both spin manifolds. The main contribution to the fast drop of the quintet population during first few fs is due to the $(S=2,M_S=+2)\rightarrow(S=1,M_S=+1)$ transitions. Quintets with $M_S=-1$ and $-2$ start to be populated only after about 1\,fs. Because of this quintet-triplet population transfer, $\rho_{\uparrow}$ notably decreases during the first 3~fs (Fig.~\ref{Fig_soc}c)).  After about 4\,fs the system almost equilibrates, i.e.\ the 760 electronic states act like an ``electronic bath''. The corresponding populations of $M_S$ components oscillate around their mean value (see Supplement).
The spin density changes relatively slowly  from the dominating $\rho_{\uparrow}$ to the dominating $\rho_{\downarrow}$ and back due to the partial revivals of quintet's positive and negative spin projections.
The fast modulation in Fig.~\ref{Fig_soc}d) with a period of $\approx$0.32~fs can be assigned to the SOC splitting between the L$_2$ and L$_3$ bands. It is roughly the same for all interacting states and is an intrinsic property of the 2p core-hole. Thus, core-excited states demonstrate an unprecedentedly fast purely electronic spin-flip dynamics, which is two orders of magnitude faster than that driven by nuclear motion in conventional spin-crossover.~\cite{Auboeck2015}

\textit{Case II.} The spin dynamics upon excitation with a spectrally broad (width of 5.0~eV) light pulse centered at  708.4~eV  (labeled 2 in Figs.~\ref{Fig_soc}a) and b)) is shown in  Fig.~\ref{Fig_soc}e).
For this excitation condition the population of all triplet states stays below 40\% within the time period of 15~fs.
As compared with case I, most notable is the absence of the rapid oscillations. This is due to the fact that the temporal width of the pulse is longer than the 0.3~fs oscillation period dictated by SOC, i.e.\ the effect is smeared out. Further, compared to case I, there are more slowly oscillating components in Fig.~\ref{Fig_soc}e). This can be traced to the fact that the initial state before excitation is  an incoherent thermal mixture of different $M_S$ components. Hence, the pattern of $\Delta M_S=0,\pm1$ transitions, which are possible upon excitation, changes.

\begin{figure}
\includegraphics{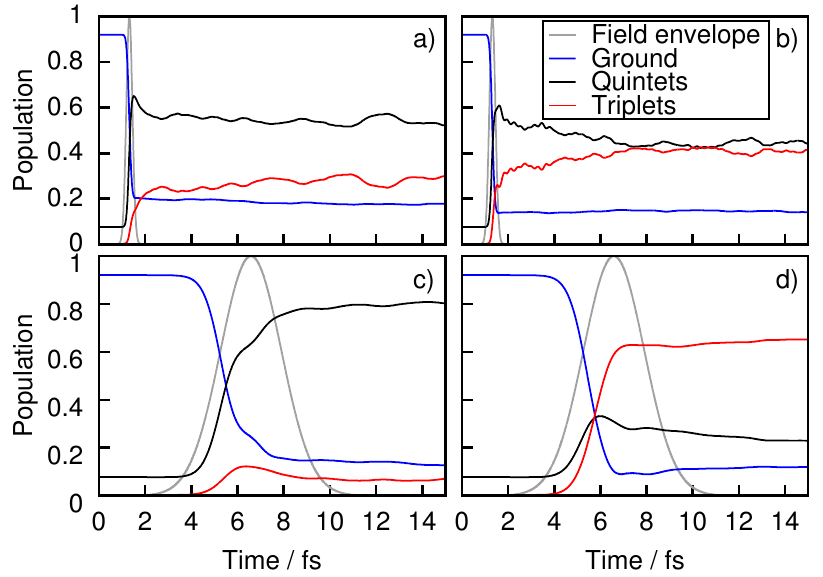}
\caption{Spin dynamics initiated by the explicit field excitation with different carrying frequencies and pulse widths corresponding to: a) 706.9~eV and 5.0~eV, b) 711.5~eV and 5.0~eV, c) 706.9~eV and 0.5 eV, d) 711.5~eV and 0.5~eV, respectively (for the spectral overlap with the SOC states, cf. Fig.~\ref{Fig_soc}b)). The pulses are centered at $t_0=$1.32~fs for a) and b) and at 6.58~fs for c) and d).  The field amplitude is $E_0=$2.5\,$\rm E_he^{-1}bohr^{-1}$ for panels a) and b) and $E_0=$1.5\,$\rm E_he^{-1}bohr^{-1}$ for c) and d).
}
\label{Fig_exc}
\end{figure}

The actual degree of quintet-triplet spin mixing is rather sensitive to the excitation conditions. This is shown in Fig.~\ref{Fig_exc}, where the spin dynamics is given for  two different excitation frequencies, 706.9 and 711.5~eV, and two pulse widths, 0.5 and 5.0~eV. Here, the excitation frequencies correspond to  spectral regions with small and notable SOC mixing (cf. the arrows in Fig.~\ref{Fig_soc}a) and the  numbered ranges in Fig.~\ref{Fig_soc}b)). The pulse amplitude has been chosen such as to yield a similar depletion of the ground state of about 80\% (green curves, Fig.~\ref{Fig_exc}). Comparing panels a) and b) with  Fig.~\ref{Fig_soc}e) one notices similar oscillations, but noticeably different quintet/triplet ratios, reflecting the spin-mixing in the excitation range. Decreasing the spectral width of the pulses (panels c) and d)) washes out the oscillations almost completely. Further, the spectral selectivity with respect to the spin-mixing becomes even more pronounced. A slight modification of the excitation frequency from 706.9~eV to 711.5~eV, changes the quintet/triplet ratio at 15~fs from 0.4 to 11.3.

Summarizing, we have studied the spin-flip dynamics, which is driven solely by SOC on a timescale where nuclear motion can be neglected. Such dynamics should be typical for states having core-holes with a nonzero orbital momentum.  This process can be considered as an elementary step of the conventional nuclear dynamics driven spin crossover,~\cite{Cannizzo2010} analogously to charge migration~\cite{Kuleff2014} being an elementary step of electron-nuclear dynamics leading to charge transfer.~\cite{Oliver} In both cases, electronic wave packet dynamics is ultimately coupled to nuclear motions, eventually leading to charge or spin localization. Using the example of a prototypical 3rd period transition metal complex, it has been demonstrated that soft X-ray light can trigger spin-dynamics, which is faster than the lifetime of the 2p core hole ($\approx$4~fs and $\approx$10~fs for Fe L$_2$ and L$_3$, respectively).~\cite{Ohno_2009} Interestingly, the actual spin-mixture can be controlled to quite some extent with modest effort, i.e.\ by small changes of pulse duration and carrier frequency.  Although the reported ultrafast spin-flip is of predominant intra-atomic character, we expect the dynamics to be influenced by the chemical environment (ligands), especially in cases where covalent ligand-metal interactions substantially change the electronic structure. 

Given the recent progress in  high harmonic generation~\cite{Dromey2006,Zhang2015,Popmintchev2015,Teichmann2016} and free-electron lasers~\cite{Picon_2016} the experimental verification of the electronic spin-flip process and its use for manipulating spin dynamics appears to be within reach.

\begin{acknowledgments}
We acknowledge financial support by the Deanship of Scientific Research (DSR), King Abdulaziz University, Jeddah, (grant No.\ D-003-435).
\end{acknowledgments}


%

\end{document}